\documentclass[11pt,reqno]{amsart}
\usepackage{indentfirst, amssymb,amsmath,amsthm,hyperref}
\usepackage{physics,breqn}
\usepackage[mathscr]{euscript}

\usepackage[caption=false]{subfig}
\usepackage{graphicx}
\usepackage{tabulary}
\usepackage{epstopdf}

\newtheorem{theo}{Theorem}

\newtheorem{rem}{Remark}
\newtheorem{pro}{Proposition}
\newtheorem{cor}{Corollary}

\newcommand{\be}{\begin{equation}}
	\newcommand{\ee}{\end{equation}}
\newcommand{\beas}{\begin{eqnarray*}}
	\newcommand{\eeas}{\end{eqnarray*}}
\newcommand{\bea}{\begin{eqnarray}}
	\newcommand{\eea}{\end{eqnarray}}

\numberwithin{equation}{section}

\begin{document}
	
\setlength{\unitlength}{1mm} \baselineskip .45cm
\setcounter{page}{1}
\pagenumbering{arabic}
	
\title[ projective curvature tensor\ldots]
{Impact of projective curvature tensor in $f\left(R,G\right)$, $f\left(R,T\right)$ and $f\left(R,L_{m}\right)$-gravity}
	
\author[ ]
{Young Jin Suh$^{1}$, Krishnendu De *$^{\,2}$ and Uday Chand De$^{3}$}

\address{$^{1}$  Department of Mathematics and RIRCM,
 Kyungpook National University,
Daegu-41566, South Korea}
\email{yjsuh@knu.ac.kr}
\address{$^2$ Department of Mathematics,
 Kabi Sukanta Mahavidyalaya,
The University of Burdwan.
Bhadreswar, P.O.-Angus, Hooghly,
Pin 712221, West Bengal, India. ORCID iD: https://orcid.org/0000-0001-6520-4520}
\email{krishnendu.de@outlook.in }

\address{$^3$Department of Pure Mathematics, University of Calcutta, 35 Ballygunge Circular Road, Kolkata 700019, West Bengal, India. ORCID iD: https://orcid.org/0000-0002-8990-4609}
\email {uc$_{-}$de@yahoo.com, ucde1950@gmail.com}

\begin{abstract}
This article concerns with the characterization of a spacetime and modified gravity, such as $f\left(R,G\right)$, $f\left(R,T\right)$ and $f\left(R,L_{m}\right)$-gravity equipped with the projective curvature tensor. We establish that a projectively flat perfect fluid spacetime represents dark energy era. Also, we prove that a projectively flat perfect fluid spacetime is either locally isometric to Minkowski spacetime or a de-Sitter spacetime. Furthermore, it is shown that a perfect fluid spacetime permitting harmonic projective curvature tensor becomes a generalized Robertson-Walker spacetime and is of Petrov type $I$, $D$ or $O$. Lastly, we investigate the effect of projectively flat perfect fluid spacetime solutions in $f\left(R,G\right)$, $f\left(R,T\right)$ and $f\left(R,L_{m}\right)$-gravity, respectively. We also investigate the spacetime as a $f\left(R,G\right)$-gravity solution of and use the flat Friedmann-Robertson-Walker metric to establish a relation among jerk, snap, and deceleration parameters. Numerous energy conditions are studied in terms of Ricci scalar with the model $f\left(R,G\right)=\exp(R)+\alpha \left(6G\right)^{\beta}$. For this model, the strong energy condition is violated but the weak, dominant and null energy conditions are fulfilled, which is in excellent accordance with current observational investigations that show the universe is now accelerating.

\end{abstract}
	
\keywords{perfect fluid spacetime, Robertson-Walker spacetime, projective curvature tensor,  energy condition,  modified gravity. \\
*Corresponding author\\
The first author was supported by the grant NRF-2018-R1D1A1B-05040381 from National Research Foundation of Korea.}

\maketitle

\section{Introduction}
In general relativity (briefly, GR) theory, a spacetime is a Lorentzian manifold $M^{4}$ with the metric (Lorentzian) $g$ of signature $\left(+,+,+,-\right)$ which admits a globally time-oriented vector. As a spontaneous source of Einstein's field equations (briefly, EFEs) that are compatible with the Bianchi identities, perfect fluids (briefly, PFs) play an outstanding role in the theory of GR. In several sectors of physics, including nuclear physics, plasma physics, and astrophysics, relativistic PF models are of immense interest.  Numerous researchers have studied spacetimes in several techniques in (\cite{B20}, \cite{kde1}, \cite{kde2}, \cite{GD16}).\par

A Lorentzian manifold $M^{n}$ $\left(n\geq4\right)$ whose metric can be expressed as
\begin{equation}\label{1.1}
	ds^{2}=-\left(d\zeta\right)^{2}+\phi^{2}\left(\zeta\right)g^{\ast}_{v_{1}v_{2}}dx^{v_{1}}dx^{v_{2}},
\end{equation}
in which $\phi$ is a function of $\zeta$  and $g^{\ast}_{v_{1}v_{2}}=g^{\ast}_{v_{1}v_{2}}\left(x^{v_{3}}\right)$ are only functions of $x^{v_{3}}$  $(v_{1},v_{2},v_{3}$$=2,3,\ldots,n)$ is named a generalized Robertson Walker (briefly, GRW) spacetime (\cite{ARS95}, \cite{bychen}, \cite{C14}). Equation \eqref{1.1} can also be shaped as the warped product $-\mathcal{I}\times\phi^{2}\tilde{M}$, in which $\tilde{M}$ denotes an $\left(n-1\right)$-dimensional Riemannian manifold and the open interval $\mathcal{I}$ is contained in $\mathbb{R}$. If $M$ is of constant sectional curvature and of dimension three, this GRW spacetime turns into a Robertson Walker (briefly, RW) spacetime.\par

The spacetime $M^{4}$ is described as a PF- spacetime if the Ricci tensor $R_{lk}$ obeys
\begin{equation}\label{1.2}
	R_{lk}=cg_{lk}+du_{l}u_{k},
\end{equation}
in which $c$, $d$ indicate scalars and $u_{k}$ stands for a unit time-like vector ($u_{k}u^{k}=-1$) also known as a flow vector or velocity vector. According to GR theory, the matter field is indicated by $T_{lk}$ (symmetric tensor field), named the energy-momentum tensor (briefly, EMT) and due to the absence of the heat conduction term, the fluid is called perfect \cite{HE73}. For a PF-spacetime, the EMT \cite{O83} is of the shape
\begin{equation}\label{1.3}
	T_{lk}=p g_{lk}+\left(p+\mu\right)u_{l}u_{k},
\end{equation}
in which $\mu$ and $p$ indicate energy density and isotropic pressure respectively. Additionally, an equation of state (briefly, EoS) having the shape $p=p\left(\mu\right)$ links $p$ and $\mu$, and the PF-spacetime is named as isentropic. Furthermore, this spacetime is known as stiff matter for $p=\mu$. According to \cite{C15}, the PF-spacetime is represented the radiation era if $p=\dfrac{\mu}{3}$, the dust matter fluid if $p=0$, and the dark energy period if $p+\mu=0$.\par

According to the EFEs for a gravitational constant $\kappa$,
\begin{equation}\label{1.4}
	R_{lk}-\dfrac{1}{2}\,g_{lk}R=\kappa T_{lk},
\end{equation}
in which $R$ denotes the Ricci scalar. The equation (\ref{1.2})can be obtained from the equations (\ref{1.3}) and (\ref{1.4}) \cite{manticamolinaride}.\par

From the perspective of differential geometry, the projective curvature tensor $P$ is a significant tensor and  $P$ vanishes if and only if the manifold $M^{n}$ ($n\geq 3$) is locally projectively flat. This time, $P$ is described by \cite{mrs}
\begin{equation}\label{1.6}	P_{lijk}=R_{lijk}-\dfrac{1}{\left(n-1\right)}\left\{g_{lk}R_{ij}-g_{lj}R_{ik}\right\},
\end{equation}
in which $R_{lijk}$ stands for the curvature tensor. According to (pp. 84-85 of \cite{yb}), a manifold is projectively flat if and only if it has constant curvature.\par

On the other hand, Weyl tensor performs a significant role in both geometry and relativity theory. Several researchers have characterized spacetimes with Weyl tensor. The Weyl tensor $C$ is defined by
\begin{eqnarray}\label{conf}
  C_{hijk} &=&  R_{hijk}+\frac{R}{(n-1)(n-2)}\{g_{hk}g_{ij}-g_{hj}g_{ik}\}\nonumber\\&&
  -\frac{1}{n-2}(g_{ij}R_{hk}-g_{ik}R_{hj}+g_{hk}R_{ij}-g_{hj}R_{ik})
  ,
\end{eqnarray}
where $R_{hijk}$ stands for the curvature tensor.\par
Moreover, we know that
\begin{eqnarray}\label{6.4}
	\nabla_{k}C^{k}_{lij} &=& \frac{1}{2}[\{\nabla_{j} R_{li}-\nabla_{i} R_{lj}\} - \frac{1}{2(n-1)}\{g_{li}\nabla_{j}R-g_{lj}\nabla_{i}R\}].
\end{eqnarray}
The Weyl tensor is called harmonic if $\nabla_{k}C^{k}_{lij} =0$. The harmonicity of the tensor appears in conservation laws of physics.\par

It is widely circulated that in GR, energy conditions (briefly, ECs) are essential tools to investigate black holes and wormholes in many modified gravities (\cite{BIBY17}, \cite{HE73}). In \cite{RBB92}, the Raychaudhuri equations  which throw back the character of gravity through $R_{jk}v^{j}v^{k}\geq0$ (the positivity condition), in which $v^{j}$ indicates a null vector, were used to produce the ECs.  In geometry, the last stated condition is equivalent to the null energy condition (briefly, NEC) $T_{jk}u^{j}u^{k}\geq0$ in GR theoty. Moreover, the weak energy condition (briefly, WEC) states that  $T_{jk}u^{j}u^{k}\geq0,$ for all time-like vector $u^{j}$ and presumes a local energy density which is positive. Also, we know that \cite{DS99} a spacetime fulfills the strong energy condition (briefly, SEC) if for all time-like vectors $u$, $R_{hj}v^{h}v^{j}\geq0$ holds. Various modifications to EFE have been developed and thoroughly investigated for modified gravity theories in (see \cite{CDTT04, NO03, PB06}).\par

The theory of $f(R,G)$-gravity was one of these modified theories \cite{EMOS10}, that was created by replacing the original Ricci scalar $R$ with a function of  $R$ and $G$. In their investigation of the stability of the power-law and de-Sitter solutions in the $f(R,G)$ theory, the authors of \cite{DSG12} found that both depend on the structure of the $f(R,G)$-gravity and the parameters of the model, and that gravitational action strongly contributes to the stability of the solutions. The weak-field limit of $f(R,G)$ by choosing the parametrized Post-Newtonian formalism was explored in \cite{LLR14}. The probability to get inflation by taking into account a general  $f(R,G)$ theory was demonstrated in \cite{LPC15}. Also in \cite{AD14}, the authors considered the implementation of ECs for flat Friedmann cosmological models and analysed them in relation to Hubble, deceleration, snap, and jerk parameters in $f\left(R,G\right)$-gravity.\par

Moreover, $f\left(R\right)$-gravity is generalized by $f\left(R,T\right)$-gravity. In \cite{HLNO11}, Harko et al. were the first to present this modified theory of gravity. Ordines et al.\cite{ord} have mentioned the modifications in Earth's atmospheric models with the use of the $f\left(R,T\right)$-gravity. Many authors have investigated $f\left(R,T\right)$-gravity features from various angles (see \cite{kde}, \cite{sin}).\par

Also, another modified theory was the $f\left(R,L_{m}\right)$-gravity, introduced in \cite{HL10} by Harko and Lobo. It is a spontaneous generalization of $f\left(R\right)$-gravity (\cite{BBHL07}, \cite{CNO18}, \cite{CNOT06}) that directly links any arbitrary function of $R$ with the matter-related Lagrangian density $L_{m}$. The application of this gravity to the situation of arbitrary coupling among matter and geometry was done in \cite{H08}. Some specific models, for instance, $f\left(R,L_{m}\right)=\lambda+\dfrac{R}{2}+L_{m}$ ($\lambda>0$ is an arbitrary constant), suggested by Harko and Lobo \cite{HL10}.\par
In \cite{LDMS21}, the authors have investigated the $f\left(R\right)$-gravity in a projectively flat spacetime and  analyze their outcomes utilizing two common models of $f\left(R\right)$-gravity.\par

These findings served as an inspiration for the present article, which is designed to examine ECs in term of Ricci scalar $R$ in a projectively flat PF-spacetime solutions fulfilling $f\left(R,G\right)$, $f\left(R,T\right)$ and $f\left(R,L_{m}\right)$-gravity respectively and we set a new model $f\left(R,G\right)=\exp(R)+\alpha \left(6G\right)^{\beta}$ ($\alpha$ is scalar) to explain ECs. \par

After preliminaries in Section $3,$ the properties of PF-spacetime permitting projective curvature tensor are explored. Finally, we provide projectively flat PF-spacetime solutions in $f\left(R,G\right)$, $f\left(R,T\right)$ and $f\left(R,L_{m}\right)$-gravity, respectively in the last three Sections.

\section{Preliminaries}
We choose throughout the article a spacetime of dimension $4$. If $P=0$ at each point of the spacetime, then the spacetime is named projectively flat. At first we consider projectively flat spacetime. Then the equation \eqref{1.6} yields
\begin{equation}\label{2.1}
	R_{lijk}=\dfrac{1}{3}\left\{g_{lk}R_{ij}-g_{lj}R_{ik}\right\},
\end{equation}
from which we can get
\begin{equation}\label{2.1a}
	R_{iljk}=\dfrac{1}{3}\left\{g_{ik}R_{lj}-g_{ij}R_{lk}\right\}.
\end{equation}
Since $R_{lijk}+R_{iljk}=0$, hence the foregoing two equations give
\begin{equation}\label{2.1b}
	\left\{g_{lk}R_{ij}-g_{lj}R_{ik}+g_{ik}R_{lj}-g_{ij}R_{lk}\right\}=0.
\end{equation}

Multiplying \eqref{2.1b} by $g^{ij}$, we acquire
\begin{equation}\label{2.2}
	R_{lk}=\dfrac{R}{4}\,g_{lk}.
\end{equation}
Therefore, we write:
\begin{pro}
	A projectively flat spacetime represents an Einstein spacetime.
\end{pro}
\noindent
Making use of \eqref{2.2} in \eqref{2.1}, we have
\begin{equation}\label{2.3}
	R_{lijk}=\dfrac{R}{12}\left\{g_{ij}g_{lk}-g_{ik}g_{lj}\right\}.
\end{equation}
Hence, the spacetime is of constant sectional curvature.\par
Let the space be a space of constant curvature $\lambda$. Then, we acquire
\begin{equation}\label{2.3a}
	R_{lijk}=\lambda \left\{g_{ij}g_{lk}-g_{ik}g_{lj}\right\},
\end{equation}
which entails $R_{lk}=3\lambda g_{lk}$. Substituting this value in \eqref{1.6}, we infer $P_{lijk}=0$.\par
Hence, we state:
\begin{pro}
	A spacetime is projectively flat if and only if the spacetime is of constant sectional curvature.
\end{pro}

\begin{rem}
  A space of constant curvature is an Einstein space, but generally, the converse is not valid. Although, an Einstein space of dimension three is a space of constant curvature. As for example ` a space with Schwarzschild metric is an Einstein space, but not a space of constant curvature.'
\end{rem}
\begin{rem}
The foregoing spacetime is either a de-Sitter or anti de-Sitter spacetime.
\end{rem}

For $n=4$, taking covariant derivative of \eqref{1.6}, we acquire
\begin{equation}\label{2.4}
	\nabla_{h}P^{h}_{ijk}=\nabla_{h}R^{h}_{ijk}
-\dfrac{1}{3}\left\{\delta^{h}_{k}\nabla_{h}R_{ij}-\delta^{h}_{j}\nabla_{h}R_{ik}\right\}.
\end{equation}
It is well-known that
\begin{equation}\label{2.5}
	\nabla_{h}R^{h}_{ijk}=\nabla_{k}R_{ij}-\nabla_{j}R_{ik}.
\end{equation}

Using \eqref{2.5} in \eqref{2.4} provides
\begin{equation}\label{2.7}
	\nabla_{h}P^{h}_{ijk}=\dfrac{2}{3}\left\{\nabla_{k}R_{ij}-\nabla_{j}R_{ik}\right\}.
\end{equation}
If the tensor $P^{h}_{ijk}$ is harmonic, that is, $\nabla_{h}P^{h}_{ijk}=0$, then \eqref{2.7} gives
\begin{equation}\label{2.11}
	\nabla_{k}R_{ij}-\nabla_{j}R_{ik}=0,
\end{equation}
which means that $R_{lk}$ is of Codazzi type.\\
Conversely, if $R_{lk}$ is of Codazzi type, then
\begin{equation}\label{2.12}
	\nabla_{k}R_{ij}-\nabla_{j}R_{ik}=0.
\end{equation}

Hence, the equation \eqref{2.7} turns into
\begin{equation}\label{2.14}
	\nabla_{h}P^{h}_{ijk}=0.
\end{equation}
This means that projective curvature tensor is harmonic. So we have
\begin{pro}
	In a semi-Riemannian space 
$P$ is harmonic if and only if $R_{lk}$ is of Codazzi type.
\end{pro}

\section{PF-spacetime permitting projective curvature tensor}
\noindent
Now consider a projectively flat PF-spacetime obeying EFE.\par
From \eqref{1.3}, \eqref{1.4} and \eqref{2.2}, we acquire	
\begin{equation}\label{3.1}
	\left(\kappa p+\dfrac{R}{4}\right)g_{lk}+\kappa\left(p+\mu\right)u_{l}u_{k}=0.
\end{equation}
Multiplying \eqref{3.1} with $g^{lk}$ implies that
\begin{equation}\label{3.2}
	3\kappa p+R-\kappa\mu=0.
\end{equation}
Also, multiplying \eqref{3.1} with $u^{l}$ gives
\begin{equation}\label{3.3}
	R=4\kappa\mu.
\end{equation}
Combining the equations \eqref{3.2} and \eqref{3.3} yield
\begin{equation}\label{3.4}
	p+\mu=0,
\end{equation}
which entails a dark energy era \cite{C15}. Therefore, we write:
\begin{theo}
A projectively flat PF-spacetime obeying EFE becomes a dark energy era.
\end{theo}
\noindent
Now, equations \eqref{1.3} and \eqref{1.4} together provide
\begin{equation}\label{3.5}
	R_{lk}=\left(\kappa p+\dfrac{R}{2}\right)g_{lk}+\kappa\left(p+\mu\right)u_{l}u_{k}.
\end{equation}
Multiplying \eqref{3.5} by $u^{l}u^{k}$, we obtain
\begin{equation}\label{3.6}
	R_{lk}u^{l}u^{k}=-\dfrac{R}{2}+\kappa\mu.
\end{equation}
Hence from equations \eqref{3.3} and \eqref{3.6}, we find
\begin{equation}\label{3.7}
	R_{lk}u^{l}u^{k}=-\kappa\mu.
\end{equation}
Here, we consider the spacetime under consideration meets the SEC. Then
\begin{equation}\label{3.8}
	\kappa\mu\leq0.
\end{equation}
As $\kappa>0$ and $\mu$ is non-negative, the equations \eqref{3.3} and \eqref{3.8} provide us
\begin{equation}\label{3.9}
	R=0.
\end{equation}
Then \eqref{2.3} infers $R_{lijk}=0$, which means that the spacetime has zero sectional curvature. Therefore a projectively flat PF-spacetime and Minkowski spacetime are locally isometric (\cite{DS99}, p. 67).\par

Therefore, we state:
\begin{theo}
	A projectively flat $\mathrm{PF-spacetime}$ fulfilling the $\mathrm{SEC}$ and a Minkowski spacetime are locally isometric.
\end{theo}

As $\mu$ is non-negative, \eqref{3.3} reflects that
\begin{equation}\label{3.10}
	R\geq0,
\end{equation}
which implies $R>0$ or, $R=0$.

{\bf Case (i).} For $R=0,$ \eqref{2.3} infers $R_{lijk}=0$. Hence, this spacetime and the Minkowski spacetime are locally isometric.\par

{\bf Case (ii).} For $R>0,$ \eqref{2.3} infers that the space is of positive constant curvature. Therefore, the space is a de-Sitter spacetime \cite{DS99}.\par

Thus, we write:
\begin{theo}
	A projectively flat PF-spacetime fulfilling the $\mathrm{SEC}$ is either locally isometric to Minkowski spacetime or a de-Sitter spacetime.
\end{theo}

It is well circulated that the de-Sitter spacetime is always conformally flat. Hence, the spacetime belongs to Petrov classification $O.$\par
 Thus, we have
\begin{cor}
	A projectively flat PF-spacetime fulfilling the $\mathrm{SEC}$ belongs to Petrov classification $O$ or locally isometric to Minkowski spacetime.
\end{cor}

Next we consider the harmonic projective curvature, that is, $\nabla_{h}P^{h}_{ijk}=0$.\par

A Yang pure space \cite{GN98} is a Lorentzian manifold whose metric obeys the Yang's equation:
\begin{equation}\label{3.11}
	\nabla_{h}R_{lk}=\nabla_{k}R_{lh}.
\end{equation}
Therefore by Proposition $3$, we can say that a spacetime permitting $\nabla_{h}P^{h}_{ijk}=0$ is a Yang pure space.\par

Thus, we state:
\begin{theo}
A PF-spacetime permitting harmonic projective curvature tensor is a Yang pure space.
\end{theo}

Again $div P=0$ implies $div C=0$, since $\nabla_{l}R_{ij}=\nabla_{j}R_{il}$ and hence $R=$ constant.\par

In \cite{Mantica5}, Mantica et al established the subsequent:\par
{\bf{Theorem A.}} If $\nabla_m C_{jkl}^{m} =0$ and $R=$constant in a PF-spacetime,
then it is a $GRW$ spacetime.\par
Therefore, by Theorem A, we write:
\begin{theo}
  A PF-spacetime with $\nabla_{h}P^{h}_{ijk}=0$ represents a GRW spacetime.
\end{theo}
In a GRW spacetime ( p. 14, \cite{survey}), we have
\begin{equation*}
  \nabla_{h}C^{h}_{ijk}=0\Longleftrightarrow u^{h}C_{hijk}=0.
\end{equation*}
In dimension 4, $u^{h}C_{hijk}=0$ implies $C_{hijk}=0$ and hence the spacetime represents a RW spacetime.
\begin{theo}
  A PF-spacetime permitting harmonic projective curvature tensor represents a RW spacetime.
\end{theo}
Also, $C_{hijk}=0$ implies $C$ is purely electric\cite{her}. We know (\cite{ste}, p. 73)  that the spacetime is of Petrov type $I$, $D$ or $O$, since $C$ is purely electric.
\begin{theo}
  A PF-spacetime permitting harmonic projective curvature tensor is of Petrov type $I$, $D$ or $O$.
\end{theo}

\section{Projectively flat PF-spacetime solutions fulfilling $f\left(R,G\right)$-gravity}
\noindent
Now, we concentrate on a particular subclass of modified gravity known as $f(R,G)$-gravity. The expression for gravitational force is
\begin{equation}\label{4.1}
	S=\dfrac{1}{2\kappa}\int (-g) ^{\frac{1}{2}} f\left(R,G\right)d^{4}x+S_{\mathrm{mat}},
\end{equation}
$S_{\mathrm{mat}}$ indicates the matter action. The Gauss-Bonnet invariant $G$ is described by
\begin{equation}\label{4.2}
	G=R^{2}+R_{lijk}R^{lijk}-4R_{lk}R^{lk}.
\end{equation}
The action term of \eqref{4.1} provides the field equations as:
\begin{equation}\label{4.3}
	R_{ij}-\dfrac{R}{2}\,g_{ij}=\kappa T_{ij}+\Omega_{ij}=\kappa T_{ij}^{\mathrm{\,eff}},
\end{equation}
where
\begin{align}\label{4.4}
	\Omega_{ij}&=\nabla_{i}\nabla_{j}f_{R}-g_{ij}\Box f_{R}+2R\nabla_{i}\nabla_{j}f_{G}-2g_{ij}R\Box f_{G}-4R^{l}_{i}\nabla_{l}\nabla_{j}f_{G}\nonumber\\
	&-4R^{l}_{j}\nabla_{l}\nabla_{i}f_{G}+4R_{ij}\Box f_{G}+4g_{ij}R^{lk}\nabla_{l}\nabla_{k}f_{G}+4R_{ilkj}\nabla^{l}\nabla^{k}f_{G}\nonumber\\
	&-\dfrac{1}{2}\,g_{ij}\left(Rf_{R}+Gf_{G}-f\right)+\left(1-f_{R}\right)\left(R_{ij}-\dfrac{1}{2}\,g_{ij}R\right)
\end{align}
and $T_{ij}^{\mathrm{\,eff}}$ denotes the effective EMT. Notice that $f_{R}\equiv\dfrac{\partial f}{\partial R}$, $f_{G}\equiv\dfrac{\partial f}{\partial G}$ and $\Box$ represent the d’Alembert operator.\par

Despite of the complexity of the above expression, in \cite{cap1} Capozziello et al. established that in a Friedmann-Robertson-Walker space-time of dimension n, for any analytical $f\left(R,G\right)$ model of gravity, the tensor $\Omega_{ij}$ is a perfect fluid form. In \cite{gu}, it is proved that the field equations of the general $f\left(R,G\right)$ gravity theory are of the perfect fluid type. Geometric perfect fluids in $f\left(R,G\right)$ gravity is also studied in \cite{cap2}.\par

These field equations are utilized to acquire the ECs of $f(R,G)$ -gravity, and get the following:

\begin{align}\label{4.5}
	\mathrm{NEC}&\Longleftrightarrow p+\mu\geq0,\\\label{4.6}
	\mathrm{WEC}&\Longleftrightarrow\mu\geq0\quad\mathrm{and}\quad p+\mu\geq0,\\\label{4.7}
	\mathrm{DEC}&\Longleftrightarrow\mu\geq0\quad\mathrm{and}\quad \mu \pm p \geq0,\\\label{4.8}
	\mathrm{SEC}&\Longleftrightarrow 3p+\mu \geq0\quad\mathrm{and}\quad\mu+p\geq0,
\end{align}
where DEC indicates the dominant energy condition.\\
From \eqref{2.2}, it follows that
\begin{equation}\label{4.9}
	R^{lk}=\dfrac{R}{4}\,g^{lk}.
\end{equation}
Equations \eqref{2.2} and \eqref{4.9} together imply
\begin{equation}\label{4.10}
	R_{lk}R^{lk}=\dfrac{R^{2}}{4}\,.
\end{equation}
From \eqref{2.3}, it follows that
\begin{equation}\label{4.11}
	R^{lijk}=\dfrac{R}{12}\left\{g^{ij}g^{lk}-g^{ik}g^{lj}\right\}.
\end{equation}
Multiplying \eqref{2.3} and \eqref{4.11}, one infers
\begin{equation}\label{4.12}
	R_{lijk}R^{lijk}=\dfrac{R^{2}}{6}\,.
\end{equation}
Equations \eqref{4.2}, \eqref{4.10} and \eqref{4.12} reflect that 
\begin{equation}\label{4.13}
	G=\dfrac{R^{2}}{6}\,.
\end{equation}
Since $R$ is constant for a projectively flat spacetime, the equation \eqref{4.4} becomes
\begin{equation}\label{4.14}
	\Omega_{ij}=R_{ij}+\left(\dfrac{f}{2}-\dfrac{R}{2}\right)g_{ij}.
\end{equation}
For a PF-spacetime the EMT is given by
\begin{equation}\label{4.15}
	T_{lk}=p g_{lk}+\left(\mu+p\right)u_{l}u_{k}
\end{equation}
and
\begin{equation}\label{4.16}
	T_{lk}^{\mathrm{\,eff}}=p^{\mathrm{\,eff}}g_{lk}+\left(\mu^{\mathrm{\,eff}}+p^{\mathrm{\,eff}}\right)u_{l}u_{k},
\end{equation}
in which $p^{\mathrm{\,eff}}$ and $\mu^{\mathrm{\,eff}}$ stand for the effective isotropic pressure and energy density, respectively.\par 

Using \eqref{4.14} and \eqref{1.3} in \eqref{4.3}, we obtain
\begin{equation}\label{4.17}
	\left(\kappa p+\dfrac{f}{2}\right)g_{ij}+\kappa\left(p+\mu\right)u_{i}u_{j}=0.
\end{equation}
Multiplying \eqref{4.17} with $u^{i}$, we have
\begin{equation}\label{4.18}
	\mu=\dfrac{f}{2\kappa}\,.
\end{equation}
Again multiplying \eqref{4.17} with $g^{ij}$ and using \eqref{4.18}, we arrive at
\begin{equation}\label{4.19}
	p=-\dfrac{f}{2\kappa}\,.
\end{equation}
Hence, we provide:
\begin{theo}
	For a projectively flat $\mathrm{PF-spacetime}$ solutions satisfying $f\left(R,G\right)$-gravity, $\mu$ and $p$ are described by \eqref{4.18} and \eqref{4.19}, respectively.
\end{theo}
\noindent
The combination of \eqref{4.18} and \eqref{4.19} give
\begin{equation*}
	p+\mu=0,
\end{equation*}
which means that NEC is satisfied. When acting on neighbouring particles that are also travelling null geodesics, locally gravity is typically appealing (preferably not repulsive), which is the physical reason for NEC \cite{LDMS21}.\par
Utilizing \eqref{4.14}--\eqref{4.16} in \eqref{4.3}, one infers
\begin{equation}\label{4.20}
	R_{lk}+\left(\kappa p+\dfrac{f}{2}-\dfrac{R}{2}\right)g_{lk}+\kappa\left(\mu+p\right)u_{l}u_{k}=\kappa p^{\mathrm{\,eff}}g_{lk}+\kappa\left(\mu^{\mathrm{\,eff}}+p^{\mathrm{\,eff}}\right)u_{l}u_{k}.
\end{equation}
In light of \eqref{2.2} and \eqref{4.18}--\eqref{4.20}, we acquire
\begin{equation}\label{4.21}
	\left(\kappa p^{\mathrm{\,eff}}+\dfrac{R}{4}\right)g_{lk}+\kappa\left(\mu^{\mathrm{\,eff}}+p^{\mathrm{\,eff}}\right)u_{l}u_{k}=0.
\end{equation}
Multiplying \eqref{4.21} with $u^{l}$, we have
\begin{equation}\label{4.22}
	\mu^{\mathrm{\,eff}}=\dfrac{R}{4\kappa}\,.
\end{equation}
Again multiplying \eqref{4.21} with $g^{lk}$ and using \eqref{4.22}, we arrive at
\begin{equation}\label{4.23}
	p^{\mathrm{\,eff}}=-\dfrac{R}{4\kappa}\,.
\end{equation}

Now, we choose
\begin{equation}\label{52}
ds^{2}=a^{2}\left(t\right)\left(dx_{1}^{2}+dx_{2}^{2}+dx_{3}^{2}\right)-dt^{2},
\end{equation}
in which $a\left(t\right)$ indicates the scale factor of the universe. The preceding metric is commonly known as the flat Friedmann Robertson Walker (briefly, FRW) metric. The field equations for $f\left(R,G\right)$-gravity follow from the FRW background and a PF equation of state for ordinary matter are described by
 \begin{equation}\label{53}
    	2\dot{H_{1}}f_{R}+8H_{1}\dot{H_{1}}\dot{f_{G}}=H_{1}\dot{f_{R}}
    -\ddot{f_{R}}+4H_{1}^{3}\dot{f_{G}}-4H_{1}^{2}\ddot{f_{G}},
 \end{equation}
 \begin{equation}\label{54}
    	6H_{1}^{2}f_{R}+24H_{1}^{3}\dot{f_{G}}=f_{R}R
    -f\left(R,G\right)-6H_{1}\dot{f_{R}}+Gf_{G},
 \end{equation}
in which $H_{1}=\dfrac{\dot{a}}{a}$ indicates the Hubble parameter and overdot $ \equiv \frac{d}{dt}$. Moreover, we have
    \begin{equation}\label{55}
    	R=6\left(2H_{1}^{2}+\dot{H_{1}}\right)
    \end{equation}
    and
    \begin{equation}\label{56}
    	G=24H_{1}^{2}\left(H_{1}^{2}+\dot{H_{1}}\right).
    \end{equation}
From \eqref{4.13}, \eqref{55} and \eqref{56}, we get
    \begin{equation}\label{57}
    	H_{1}^{2}=\dfrac{R}{12}\qquad\mathrm{and}\qquad\dot{H_{1}}=0.
    \end{equation}
    As $H_{1}=\dfrac{\dot{a}}{a}$\,, $\dfrac{\dot{a}}{a}=\sqrt{\dfrac{R}{12}}$\,. Thus
    \begin{equation}\label{58}
    	\ddot{a}=\dfrac{\dot{a}^{2}}{a}\,,\qquad\dddot{a}=\dfrac{\dot{a}^{3}}{a^{2}}\qquad\mathrm{and}\qquad\ddddot{a}
    =\dfrac{\dot{a}^{4}}{a^{3}}\,.
    \end{equation}
Then, using an analogue with classical mechanics, we explain velocity, acceleration, snap and jerk in the context of cosmology. The jerk, deceleration, and snap parameters must be defined as
    \begin{equation}\label{59}
    	j=\dfrac{1}{H_{1}^{3}}\dfrac{\dddot{a}}{a}\,,\qquad q=-\dfrac{1}{H_{1}^{2}}\dfrac{\ddot{a}}{a}\qquad \mathrm{and}\qquad s=\dfrac{1}{H_{1}^{4}}\dfrac{\ddddot{a}}{a},
    \end{equation}
respectively. Using \eqref{58} in \eqref{59}, we obtain
    \begin{equation}\label{60}
    	s=j=-q.
    \end{equation}
Hence, for a projectively flat PF-spacetime fulfilling $f\left(R,G\right)$-gravity, the jerk, deceleration, and snap parameters are linked by \eqref{60}.\par

Now we concentrate on the ECs of a $f\left(R,G\right)$-gravity model.

\section*{{\bf A. $f\left(R,G\right)=\exp(R)+\alpha \left(6G\right)^{\beta}$}}
Here, making use of \eqref{4.13}, \eqref{4.18} and \eqref{4.19}, the energy density and pressure are expressed as
\begin{equation}\label{4.24a}
	\mu=\dfrac{\exp(R)+\alpha \left(6G\right)^{\beta}}{2\kappa}\,,
\end{equation}
\begin{equation}\label{4.25a}
	p=-\dfrac{\exp(R)+\alpha \left(6G\right)^{\beta}}{2\kappa}\,.
\end{equation}
Using \eqref{4.13} the foregoing equations reduce to
\begin{equation}\label{4.24}
	\mu=\dfrac{\exp(R)+\alpha \left(R\right)^{2\beta}}{2\kappa}\,,
\end{equation}
\begin{equation}\label{4.25}
	p=-\dfrac{\exp(R)+\alpha \left(R\right)^{2\beta}}{2\kappa}\,.
\end{equation}

The ECs for this setup can now be discussed using \eqref{4.24} and \eqref{4.25}. In this model the EoS reduces to $\omega=-1$. Therefore, the chosen model is consistent with the $\Lambda$CDM model. Obviously, in this model NEC is staisfied. Since WEC is the amalgamation of NEC and positive density, we examine the behavior of density parameter, DEC and SEC. \\
\begin{tabulary}{\linewidth}{CC}
	\includegraphics[height=0.25\textheight]{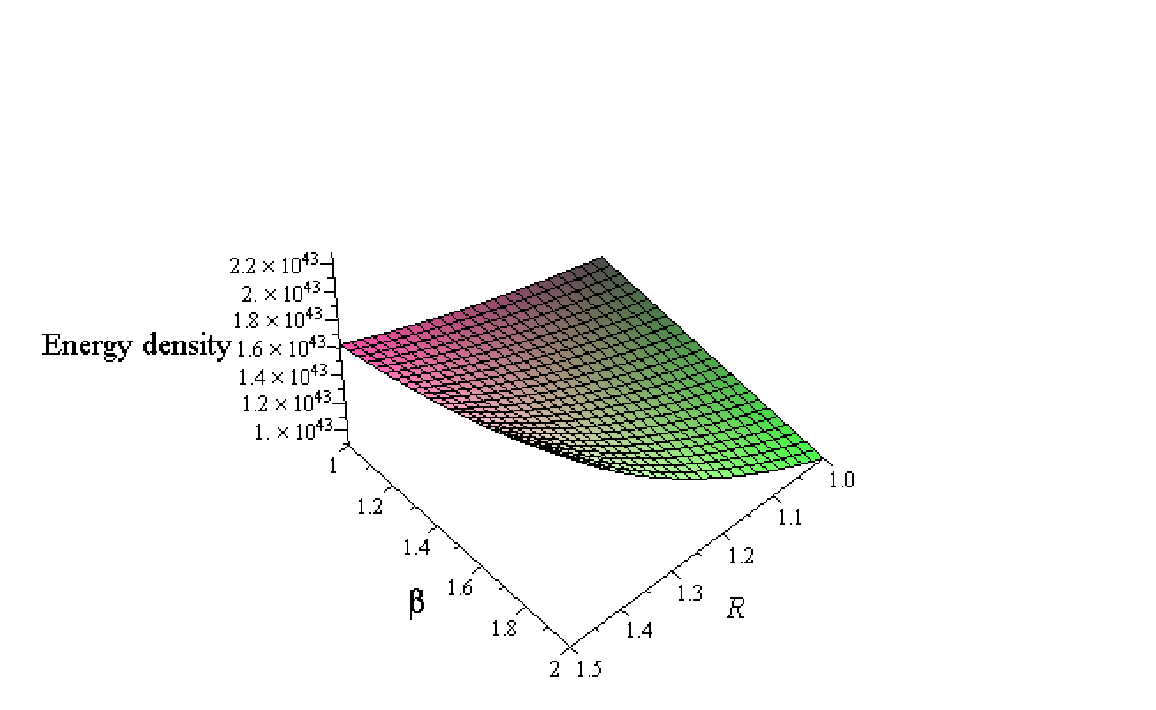}
	&
	\includegraphics[height=0.25\textheight]{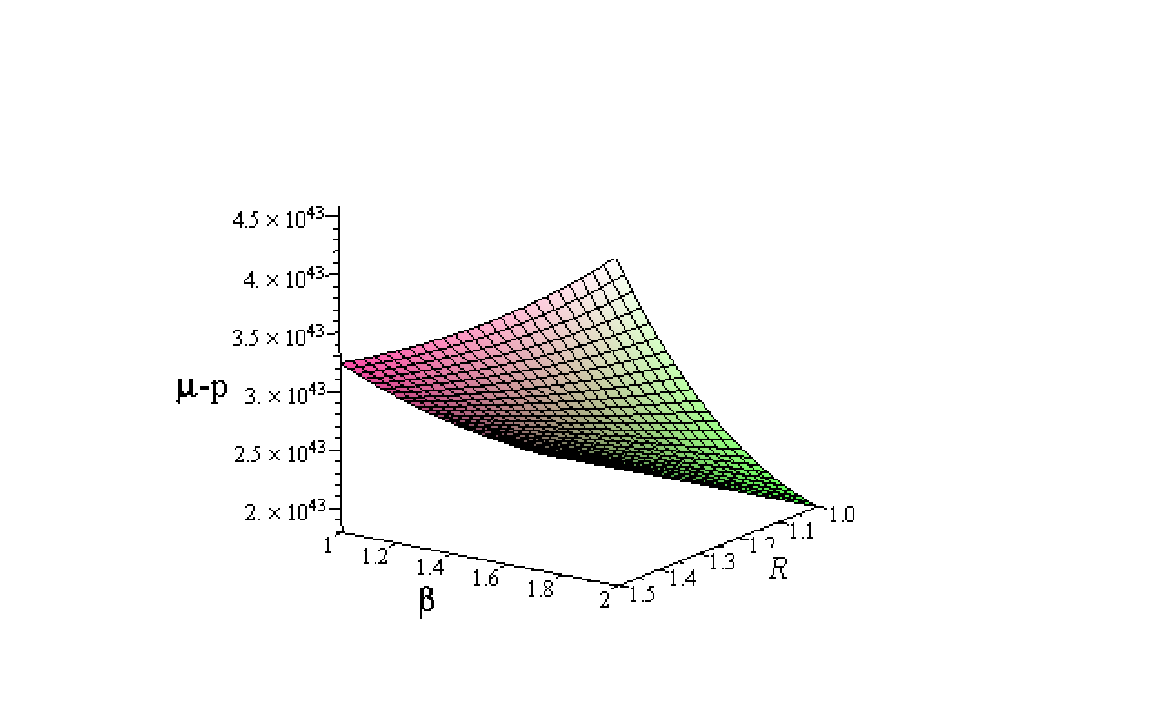}
	\\
	{\bf Fig. 1:} Development of $\mu$ with reference to $R \in [1,1.5]$ and $\beta \in [1,2]$ for $\alpha =1$ and $\kappa=2.077\times 10^{-43}$.
  &{\bf Fig. 2:} Development of $\mu-p$ with reference to $R \in [1,1.5]$ and $\beta \in [1,2]$ for $\alpha =1$ and $\kappa=2.077\times 10^{-43}$.
	
\end{tabulary}

\begin{tabulary}{\linewidth}{CC}
	
	\includegraphics[height=0.25\textheight]{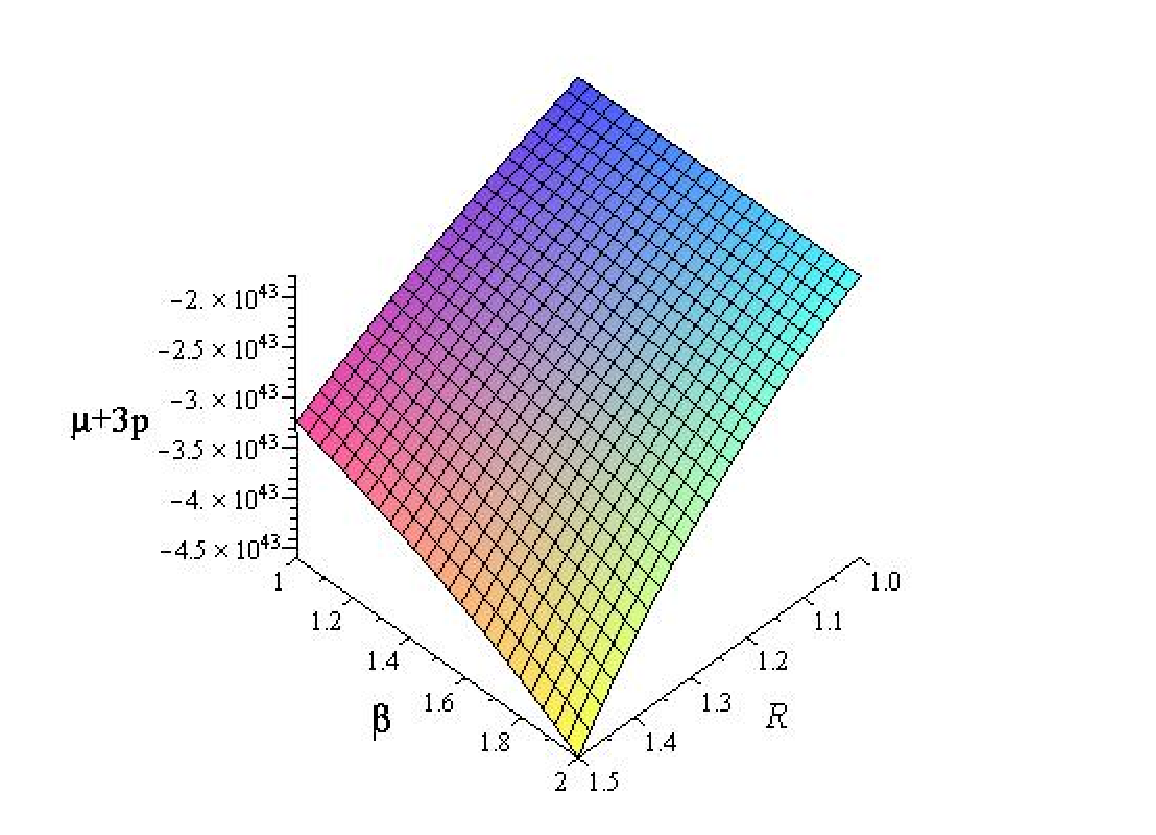}

	{\bf Fig. 3:} Development of $\mu+3p$ with reference to $R \in [1,1.5]$ and $\beta \in [1,2]$ for $\alpha =1$ and $\kappa=2.077\times 10^{-43}$.
	
\end{tabulary}
One can see from Fig. $1$, the energy density cannot be negative for the parameter ranges $R \in [1,1.5]$ and $\beta \in [1,2]$ and for greater values of $R$ and $\alpha$, it is high.  In this situation, $\mu+p$ becomes zero. As NEC belongs to WEC, as a consequence NEC and WEC are verified. Fig. $2$ gives the $\mu-p$ profile, which yields a positive value range.. Using Fig. $1$, Fig. $2$ and $\mu+p=0$, we see that DEC is verified. From Fig. $3$, we see that SEC is violated. Therefore, it describes the Universe's late-time acceleration \cite{LDMS21}.

\section{Projectively flat PF-spacetime solutions fulfilling $f\left(R,T\right)$-gravity}

The field equations of $f\left(R,T\right)$-gravity have been investigated in metric formalism for a number of special instances. Here, we set \cite{HLNO11}
\begin{equation}\label{6.1}
	f\left(R,T\right)= 2f\left(T\right)+R.
\end{equation}
Here, modified Einstein-Hilbert action term is described by
\begin{equation}\label{6.2}
	E=\int (-g)^{\frac{1}{2}}\left[\dfrac{16\pi L_{m}+f\left(R,T\right)}{16\pi}\right]\,d^{4}x,
\end{equation}
in which $L_{m}$ stands for the scalar field's matter Lagrangian. Here, the stress energy tensor is described by
\begin{equation}\label{6.3}
	T_{ij}=\dfrac{-2\delta\left(\sqrt{-g}\right)L_{m}}{\sqrt{-g}\,\delta^{ij}}\,,
\end{equation}
in which $L_{m}$ solely depends on $g$.\par
\indent
The subsequent field equations 
are acquired from  \eqref{6.2} 
\begin{align}\label{6.4}
	f_{R}\left(R,T\right)R_{ji}&-\dfrac{1}{2}f\left(R,T\right)g_{ji}
-\left[\nabla_{j}\nabla_{i}-g_{ji} \Box\right]f_{R}\left(R,T\right)\nonumber\\
	&=8\pi T_{ji}-\left[T_{ji}+\Theta_{ji}\right]f_{T}\left(R,T\right),
\end{align}
$f_{R}\left(R,T\right)$ and $f_{T}\left(R,T\right)$ are the partial derivative with regard to $R$ and $T$ respectively,  $\Box$ stands for the d’Alembert operator and
\begin{equation}\label{6.5}
	\Theta_{ji}=-2T_{ji}+g_{ji}L_{m}-2g^{lk}\dfrac{\partial^{2}L_{m}}{\partial g^{ab}\partial g^{lk}}\,.
\end{equation}
We presume that $L_{m}$ =$-p$ and utilizing \eqref{1.3}, we infer that
\begin{equation}\label{6.6}
	T_{lk}=-p g_{lk}+\left(p+\mu\right)u_{l}u_{k}.
\end{equation}
Using \eqref{6.6}, we infer the variation of stress energy as
\begin{equation}\label{6.7}
	\Theta_{lk}=-2T_{lk}-pg_{lk}.
\end{equation}
Equations \eqref{6.1} and \eqref{6.4} together produce
\begin{align}\label{6.8}
	R_{lk}=\dfrac{R}{2}g_{lk}&+8\pi T_{lk}+f\left(T\right)g_{lk}\nonumber\\
	&-2\left[T_{lk}+\Theta_{lk}\right]f^{\prime}\left(T\right).
\end{align}
The conservation of the EMT was not taken into account when the field equations were derived by Harko et al. \cite{HLNO11}. But the author of \cite{C13}, presumed the conservation of the EMT. Here we assume that the EMT is conserved in the PF-spacetime solution to the $f\left(R,T\right)$-gravity equation.\par

Making use of \eqref{6.6}, \eqref{6.7} and \eqref{6.8} provide the Ricci tensor 
as
\begin{equation}\label{6.9}	R_{lk}=\left[\dfrac{R}{2}-8p\pi+f\left(T\right)\right]g_{lk}
+\left(p+\mu\right)\left\{2f^{\prime}\left(T\right)+8\pi\right\}u_{l}u_{k}.
\end{equation}

Equations \eqref{2.2} and \eqref{6.9} reveal that
\begin{equation}\label{6.10}
	\left[\dfrac{R}{4}-8p\pi+f\left(T\right)\right]g_{lk}
+\left(p+\mu\right)\left\{2f^{\prime}\left(T\right)+8\pi\right\}u_{l}u_{k}=0.
\end{equation}
Contracting the foregoing equation reflects
\begin{equation}\label{6.11}
	R-32p\pi+4f\left(T\right)-\left(p+\mu\right)\left\{2f^{\prime}\left(T\right)+8\pi\right\}=0.
\end{equation}
Multiplying \eqref{6.10} by $g^{lk}$, we acquire
\begin{equation}\label{6.12}
	-R+32p\pi-4f\left(T\right)+4\left(p+\mu\right)\left\{2f^{\prime}\left(T\right)+8\pi\right\}=0.
\end{equation}
Adding \eqref{6.11} and \eqref{6.12}, we obtain
\begin{equation}\label{6.13}
	\left\{4\pi+f^{\prime}\left(T\right)\right\}\left(p+\mu\right)=0.
\end{equation}
From the above we conclude that \par
 either $p+\mu=0$ or, $p+\mu\neq0.$
\vskip.1in
\noindent
{\bf Case i.} If $p+\mu=0,$ then the spacetime represents dark energy era.
\vskip.1in
\noindent
{\bf Case ii.} If $p+\mu\neq0,$ then $f^{\prime}\left(T\right)+4\pi=0.$ Hence, \eqref{6.9} reveals that it is an Einstein spacetime.
\vskip.05in
\noindent
Therefore we state:
\begin{theo}
	A projectively flat spacetime satisfying $f\left(R,T\right)$-gravity represents either Einstein spacetime, or dark energy era.
\end{theo}
Equations \eqref{6.11} and \eqref{6.12} reflects
\begin{equation}\label{6.14}
	p=\dfrac{4f\left(T\right)+R}{32\pi}\,.
\end{equation}
For dust matter era ($p=0$), the previous equation turns into $f\left(T\right)=-\dfrac{R}{4}$\,. Thus, we state:
\begin{cor}
	For every viable $f\left(R,T\right)$ a projectively flat spacetime is incapable to demonstrate dust matter era.
\end{cor}

\begin{rem}
When $f\left(T\right)$ is equal to zero, $f\left(R,T\right)$-gravity transforms into $f\left(R\right)$-gravity. According to the aforementioned Theorem, in projectively flat spacetime for $f\left(R\right)$-gravity represents dark energy era. By virtue of energy density's impossibility of being negative, the EoS is $p+\mu=0,$ which means that $\lvert\mu\rvert=\lvert-p\rvert,$ that is, $\mu=\lvert p \rvert$ . Thus, a projectively flat spacetime obeys the DEC in $f\left(R\right)$-gravity. Hence, the speed of light is the fastest that matter cannot travel in a projectively flat spacetime obeying $f\left(R\right)$-gravity \cite{DS99}.
\end{rem}

\section{Projectively flat PF-spacetime solutions fulfilling $f\left(R,L_{m}\right)$-gravity}
\noindent
Here, we describe the projectively flat PF-spacetime fulfilling $f\left(R,L_{m}\right)$-gravity. According to our hypothesis, the shape of the action term is as follows:
\begin{equation}\label{5.1}
	S=\int\sqrt{-g}f\left(R,L_{m}\right)d^{4}x,
\end{equation}
 The EMT of the matter is described as
\begin{equation}\label{5.2}
	T_{lk}=-\dfrac{2}{\sqrt{-g}}\dfrac{\delta\left(\sqrt{-g}L_{m}\right)}{\delta g^{lk}}\,.
\end{equation}
Supposing that $L_{m}$ is independent of the derivatives of the metric tensor $g$. From the variation of action of \eqref{5.1}, with regard to 
$g$, the field equations of $f\left(R,L_{m}\right)$ theory are given in their modified form \cite{HL10}:
\begin{align}\label{5.3}
	f_{R}\left(R,L_{m}\right)&\left\{R_{lk}-\dfrac{R}{3}\,g_{lk}\right\}
-\dfrac{1}{6}\left\{f_{L_{m}}\left(R,L_{m}\right)L_{m}-f\left(R,L_{m}\right)\right\}g_{lk}\nonumber\\
	&=\dfrac{1}{2}\left\{T_{lk}-\dfrac{t}{3}\,g_{lk}\right\}f_{L_{m}}\left(R,L_{m}\right)
+\nabla_{l}\nabla_{k}f_{R}\left(R,L_{m}\right),
\end{align}
in which $t$ stands for the trace of the EMT. For our investigations, we take into account the model described below \cite{HL10}:
\begin{equation}\label{5.4}
	f\left(R,L_{m}\right)=\lambda+\dfrac{R}{2}+L_{m},
\end{equation}
$\lambda>0$ being an arbitrary constant. Here, assuming that the EMT has the form \eqref{1.3}, we investigate PF-spacetime solutions to the $f\left(R,L_{m}\right)$-gravity.\\
Contracting \eqref{1.3}, we obtain
\begin{equation}\label{5.5}
	t=3p-\mu.
\end{equation}
Equations \eqref{1.3}, \eqref{2.2} and \eqref{5.3}--\eqref{5.5} reflect that
\begin{equation}\label{5.6}
	\left(\dfrac{\lambda}{3}+\dfrac{R}{12}-\dfrac{\mu}{3}\right)g_{lk}-\left(p+\mu\right)u_{l}u_{k}=0.
\end{equation}
Multiplying \eqref{5.6} with $g^{lk}$ and $u^{l}$ separately, we get
\begin{equation}\label{5.7}
	\dfrac{4\lambda+R}{3}+p-\dfrac{\mu}{3}=0
\end{equation}
and
\begin{equation}\label{5.8}
	\dfrac{4\lambda+R}{12}+p+\dfrac{2\mu}{3}=0.
\end{equation}
Equations \eqref{5.7} and \eqref{5.8} together imply
\begin{equation}\label{5.9}
	\mu=\dfrac{4\lambda+R}{4}
\end{equation}
and
\begin{equation}\label{5.10}
	p=-\dfrac{4\lambda+R}{4}\,.
\end{equation}
The combination of \eqref{5.9} and \eqref{5.10} give $\mu+p=0$. Thus, we write:
\begin{theo}
	A projectively flat $\mathrm{PF-spacetime}$ solutions obeying $f\left(R,L_{m}\right)=\lambda+\dfrac{R}{2}+L_{m}$ becomes a dark energy era.
\end{theo}
\noindent
Now, we examine the ECs for the model \eqref{5.4}. Using \eqref{5.9} and \eqref{5.10}, one can now discuss about the ECs for this configuration.\\
\begin{tabulary}{\linewidth}{CC}
	\includegraphics[height=0.25\textheight]{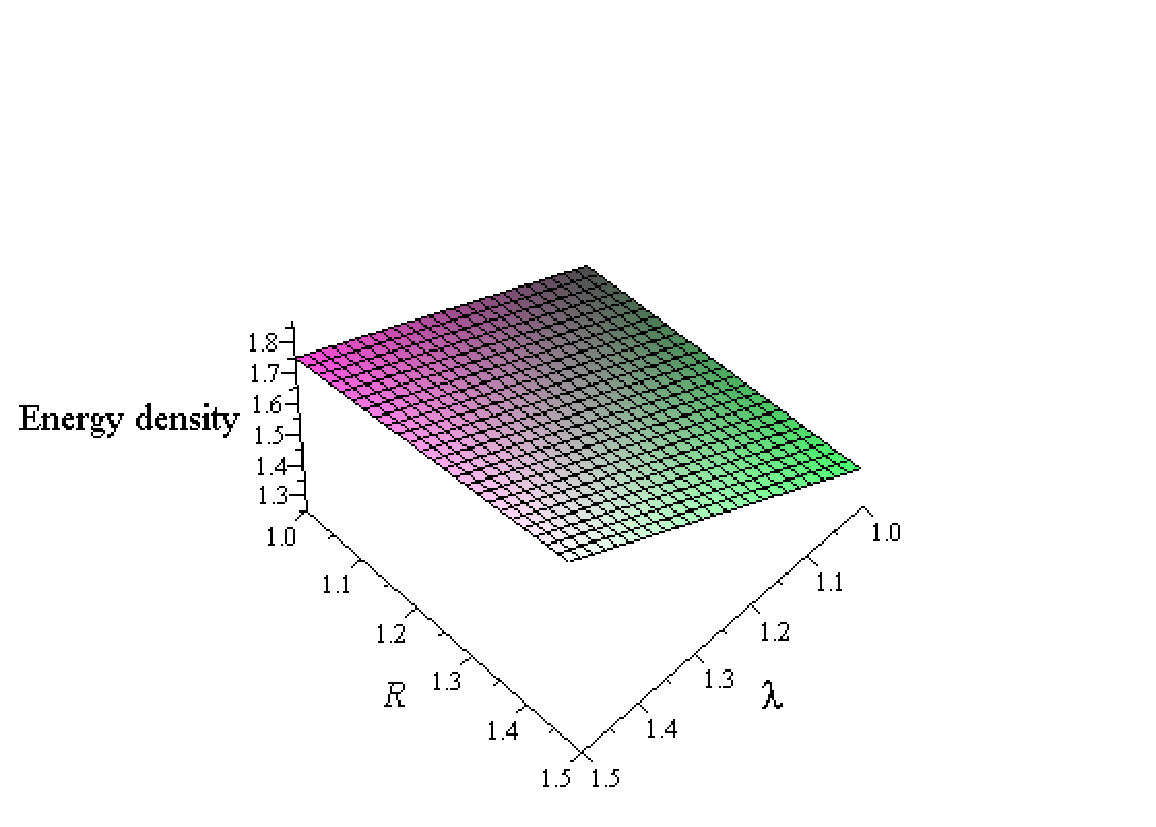}
	&
	\includegraphics[height=0.25\textheight]{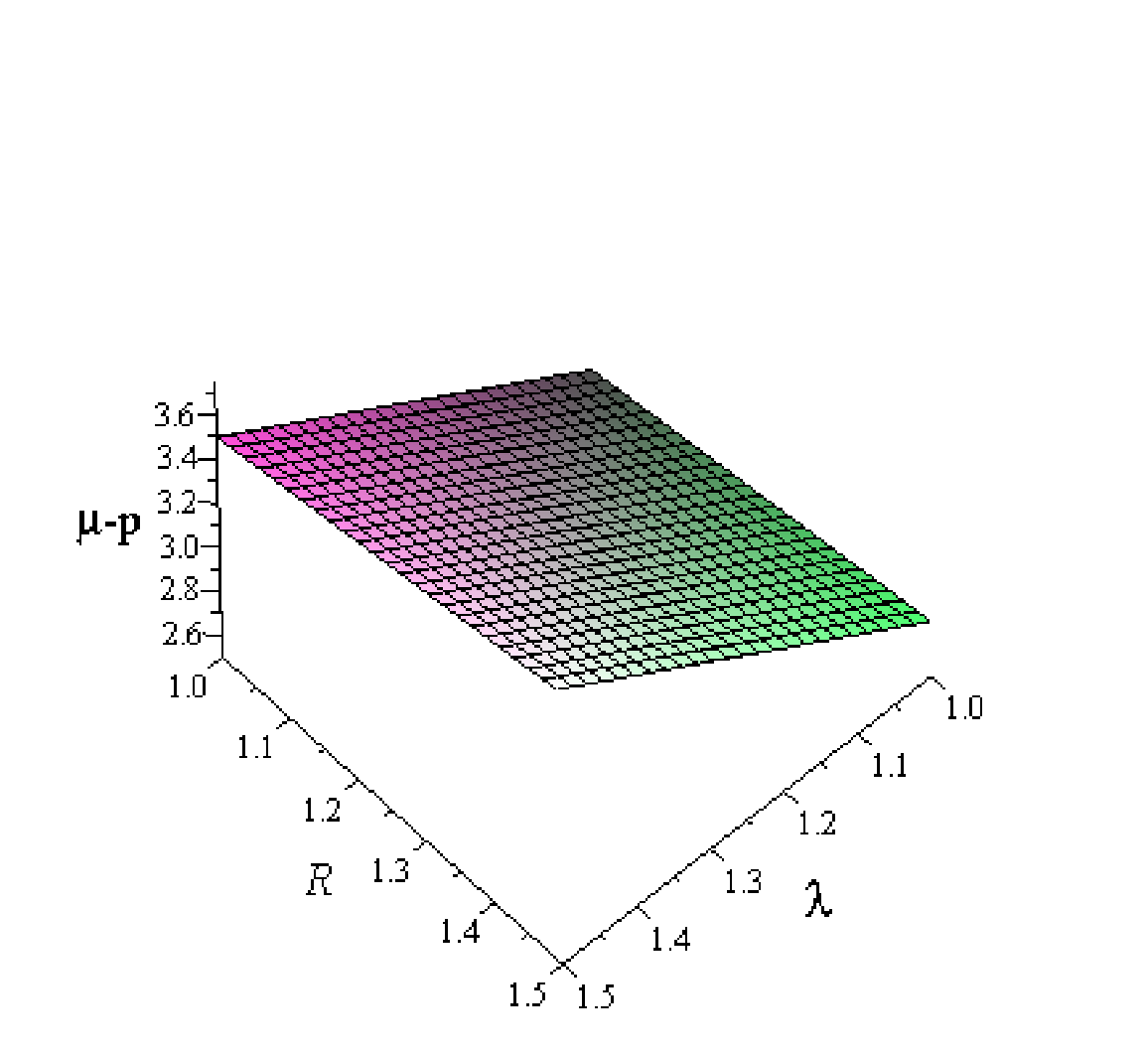}
	\\
	{\bf Fig. 4:} Development of $\mu$ with reference to  $R \in [1,1.5]$ and $\lambda \in [1,1.5]$
&{\bf Fig. 5:} Development of $\mu-p$ with reference to  $R \in [1,1.5]$ and $\lambda \in [1,1.5]$
	
\end{tabulary}
\begin{tabulary}{\linewidth}{CC}
	
	\includegraphics[height=0.25\textheight]{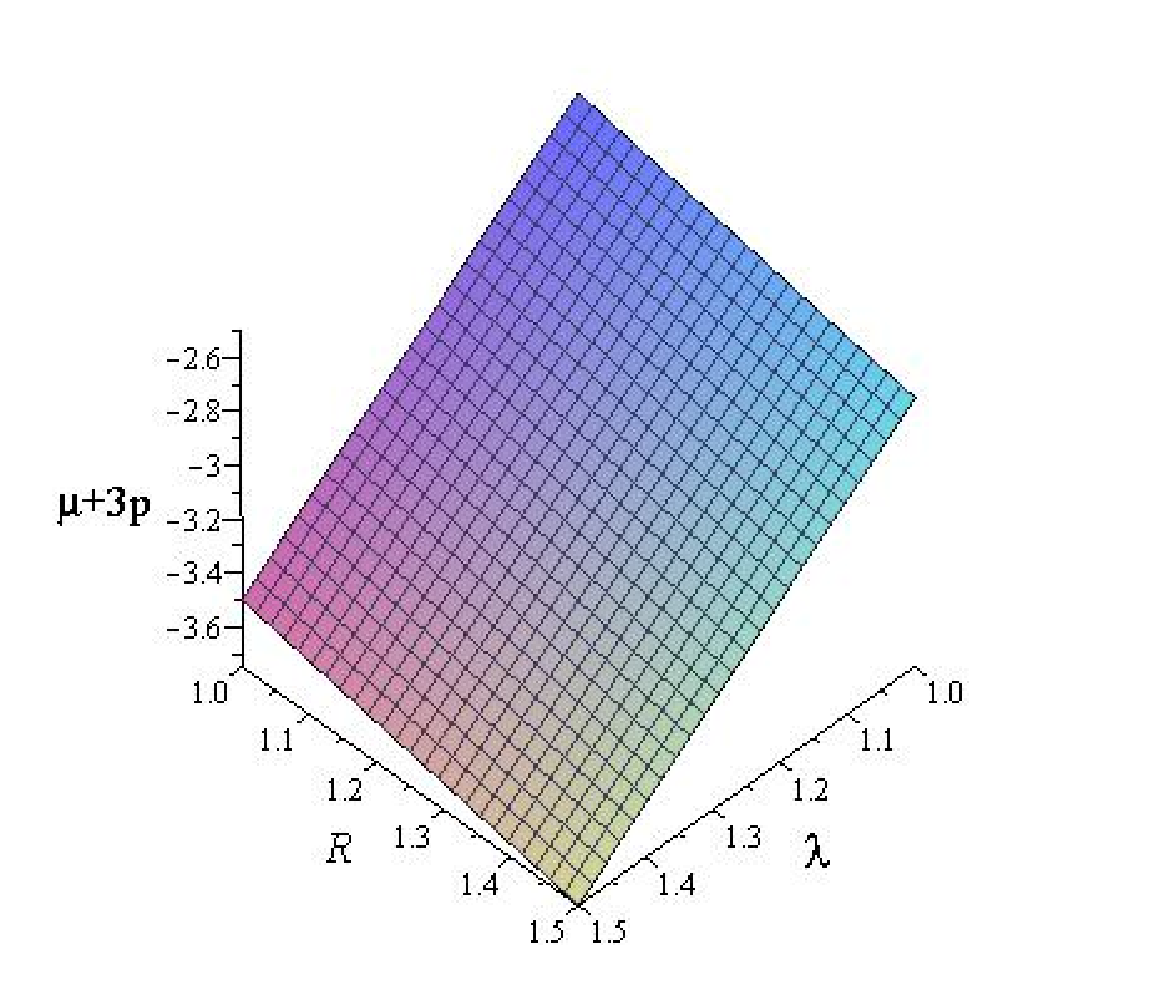}

	{\bf Fig. 6:} Development of $\mu+3p$ with reference to  $R \in [1,1.5]$ and $\lambda \in [1,1.5]$
	
\end{tabulary}
Figs. $4$, $5$, and $6$ denote the profiles of $\mu$, $\mu-p$, and $\mu+3p$. Also, for this construction, as $\mu+p=0$ , WEC and NEC are also satisfied. From the foregoing figures, we notice that DEC is verified for the parameter ranges $R \in [1,1.5]$ and $\lambda \in [1,1.5]$ but the SEC is not valid.

\section{Discussion}
Spacetime which is a time-oriented torsion-free Lorentzian manifold, serves as the foundation for the present modelling of the physical universe. According to GR theory, the appropriate EMT may be used to determine the matter content of the universe, which is agreed to behave like a PF-spacetime in cosmological models.\par

In this current investigation, we study a projectively flat PF-spacetime and show that a projectively flat PF-spacetime is either locally isometric to Minkowski spacetime or a de-Sitter spacetime. We also illustrate that a projectively flat PF-spacetime fulfilling the SEC is locally isometric to Minkowski spacetime. Furthermore, we notice that a PF-spacetime with $\nabla_{h}P^{h}_{ijk}=0$ is of Petrov type $I$, $D$ or $O$.\par

The prime focus of this article has been the exploration of projectively flat PF-spacetime solutions in relation to different modified gravity. In this article, our outcomes have been evaluated analytically and graphically. We used the analytic technique in order to construct our formulation and to assess the stability of two cosmological models, like $f\left(R,G\right)=\exp(R)+\alpha \left(6G\right)^{\beta}$ and $f\left(R,L_{m}\right)=\lambda+\dfrac{R}{2}+L_{m}$. For the first model, Figs. $1$, $2$ and $3$ show the profiles of ECs. Here we see thst although NEC, WEC and DEC were satisfied, SEC violated for this agreement. In addition, the EoS is $\dfrac{p}{\mu}=-1$, which denotes the dark energy era. However, these outcomes are consistent with the $\Lambda$CDM model. Similar to the first model, Figs. $4$, $5$ and $6$ show every ECs for the second model. The outcomes we found for the second model are also consistent with those of the first model.

\section{Declarations}
\subsection{Funding }
NA.
\subsection{Conflicts of interest/Competing interests}
The authors have no conflicts to disclose and all authors contributed equally to this work.
\subsection{Availability of data and material }
NA.
\subsection{Code availability}
NA.

\section{Acknowledgement}
We would like to thank the Referees and the Editor for reviewing the paper carefully and their valuable comments to improve the quality of the paper.


\begin{thebibliography}{00}
\bibitem{AAAABBBBBBB20}Aghanim, N., Akrami, Y., Ashdown, M., Aumont, J., Baccigalupi, C., Ballardini, M., Banday, A.J., Barreiro, R.B., Bartolo, N., Basak, S. and Battye, R., Planck $2018$ results-VI. Cosmological parameters, Astronomy \& Astrophysics, {\bf641}(2020), Article ID A6.
	
\bibitem{ARS95}Al\'ias, L., Romero, A. and S\'anchez, M., Uniqueness of complete spacelike hypersurfaces of constant mean curvature in generalized Robertson-Walker spacetimes, Gen. Relativ. Gravit., {\bf 27}(1995), 71--84.
\bibitem{AD14}Atazadeh, K. and Darabi, F., Energy conditions in $f\left(R,G\right)$-gravity, Gen. Relativ. Gravit., {\bf46}(2014), Article ID 1664.
\bibitem{BIBY17}Bamba, K., Ilyas, M., Bhatti, M.Z. and Yousaf, Z., Energy conditions in modified $f\left(G\right)$ gravity, Gen. Relativ. Gravit., {\bf49}(2017), Article ID 112.

\bibitem{BBHL07}Bertolami, O., B\"ohmer, C.G., Harko, T. and Lobo, F.S.N., Extra force in $f\left(R\right)$ modified theories of gravity, Phys. Rev. D, {\bf75}(2007), Article ID 104016.
	
\bibitem{B20}Blaga, A.M., Solitons and geometrical structures in a perfect fluid spacetime, Rocky Mt. J. Math., {\bf50}(2020), 41--53.

\bibitem{cap1}S. Capozziello, C.A. Mantica L.G. Molinari, {\it Cosmological perfect fluid in Gauss-Bonnet gravity}, Int. J. Geom. Meth. Mod. Phys. {\bf16} (09),1950133, (2019).

\bibitem{cap2} S. Capozziello, C.A. Mantica and L.G. Molinari, {\it Geometric perfect fluids from Extended Gravity}, Europhysics  Letters {\bf137} 19001 (2022).

\bibitem{CNO18}Capozziello, S., Nojiri, S. and Odintsov, S.D., The role of energy conditions in $f\left(R\right)$ cosmology, Phys. Lett. B, {\bf781}(2018), 99--106.

\bibitem{CNOT06}Capozziello, S., Nojiri, S., Odintsov, S.D. and Troisi, A., Cosmological viability of $f\left(R\right)$-gravity as an ideal fluid and its compatibility with a matter dominated phase, Phys. Lett. B, {\bf639}(2006), 135--143.

\bibitem{CDTT04}Carroll, S.M., Duvvuri, V., Trodden, M. and Turner, M.S., Is cosmic speed-up due to new gravitational physics?, Phys. Rev. D, {\bf70}(2004), Article ID 043528.
\bibitem{C13}Chakraborty, S., An alternative $f\left(\boldsymbol{r},\mathcal{T}\right)$-gravity theory and the dark energy problem, Gen. Relativ. Gravit., {\bf 45}(2013), 2039--2052.
\bibitem{C15}Chavanis, P.H., Cosmology with a stiff matter era, Phys. Rev. D, {\bf92}(2015), Article ID 103004.
\bibitem{bychen} B. Y. Chen: {\em Pseudo-Riemannian Geometry, $\delta$-invariants and Applications}, World Scientific, 2011.
\bibitem{C14}Chen, B.-Y., A simple characterization of generalized Robertson–Walker spacetimes, Gen. Relativ. Gravit., {\bf46}(2014), Article ID 1833.
\bibitem{kde} De, K. and De, U.C., {\it Investigation of generalized $\mathcal{Z}$- recurrent spacetimes and $f(\mathcal{R},T)$-gravity}, Adv. Appl. Clifford Algebras {\bf 31}, 38 (2021).
\bibitem{kde1}De, K. and De, U.C., {\it Investigations on solitons in $f(\mathcal{R})$-gravity},  Eur. Phys. J. Plus (2022) 137:180.
https://doi.org/10.1140/epjp/s13360-022-02399-y
\bibitem{kde2} De, U.C., De, K., Zengin, F.O. and Demirbag, S.A., {\it Characterizations of a spacetime of quasi-constant sectional curvature and $\mathcal{F}(\mathcal{R})$-gravity}, Fortschr. Phys. 2023, 2200201. https://doi.org/10.1002/prop.202200201

\bibitem{DSG12}De la Cruz-Dombriz, \'A. and S\'aez-G\'omez, D., {\it On the stability of the cosmological solutions in $f\left(\mathcal{R},G\right)$ gravity}, Class. Quantum Grav., {\bf 29} (2012), Article ID 245014.
\bibitem{DS99}Duggal, K.L. and Sharma, R., Symmetries of spacetimes and Riemannian manifolds, Springer New York, NY 1999.

\bibitem{EMOS10}Elizalde, E., Myrzakulov, R., Obukhov, V.V. and S\'aez-G\'omez, D., $\Lambda$CDM epoch reconstruction from $F\left(R,G\right)$ and modified Gauss–Bonnet gravities, Class. Quantum Grav., {\bf27}(2010), Article ID 095007.

\bibitem{GN98}Guilfoyle, B.S. and Nolan, B.C., Yang's gravitational theory, Gen. Relativ. Gravit., {\bf30}(1998), 473--495.

\bibitem{GD16}G\"uler, S. and Demirbağ, S.A., A study of generalized quasi-Einstein spacetimes with applications in general relativity, Int. J. Theor. Phys., {\bf55}(2016), 548--562.

\bibitem{gu} M. Gürses, Y. Heydarzade, {\it FLRW-cosmology in generic gravity theories}, Eur. Phys. J. C {\bf80} :1061 (2020).

\bibitem{HLNO11}Harko, T., Lobo, F.S.N., Nojiri, S. and Odintsov, S.D., $f\left(R,T\right)$-gravity, Phys. Rev. D, {\bf 84}(2011), Article ID 024020.

\bibitem{H08}Harko, T., Modified gravity with arbitrary coupling between matter and geometry, Phys. Lett. B, {\bf 669}(2008), 376--379.

\bibitem{HL10}Harko, T. and Lobo, F.S.N., $f\left(R,L_{m}\right)$-gravity, Eur. Phys. J. C, {\bf 70}(2010), 373--379.

\bibitem{HE73}Hawking, S.W. and Ellis, G.F.R., The Large Scale Structure of Space-Time, Cambridge University Press, London, 1973.
\bibitem{her} Hervik, S., Ortaggio, M. and Wylleman, L., {\it Minimal tensors and purely electric or magnetic spacetimes of arbitrary dimension}, Class. Quantum Grav., {\bf{30}}(2013), 165014.
\bibitem{LLR14}Laurentis, M.D. and Lopez-Revelles, A.J., Newtonian, Post-Newtonian and Parametrized Post-Newtonian limits of $f\left(\mathcal{R},G\right)$ gravity, Int. J. Geom. Methods Mod. Phys., {\bf11}(2014), Article ID 1450082.
		
\bibitem{LPC15}Laurentis, M.D., Paolella, M. and Capozziello, S., Cosmological inflation in $f\left(\mathcal{R},G\right)$ gravity, Phys. Rev. D, {\bf91}(2015), Article ID 083531.
\bibitem{LDMS21}Loo, T.-H., De, A., Mandal, S. and Sahoo, P.K., How a projectively flat geometry regulates $F\left(R\right)$-gravity theory?, Phys. Scr., {\bf96}(2021), Article ID 125034.
\bibitem{manticamolinaride} C. A. Mantica, L. G. Molinari and U. C. De, {\it A condition for a perfect fluid spacetime to be a generalized Robertson-Walker spacetime}, J. Math. Phys. {\bf57} (2) (2016), 022508.
\bibitem{survey} C. A. Mantica and L. G. Molinari, {\it Generalized Robertson Walker spacetimes-A survey}, Int. J. Geom. Meth. Mod. Phys.  {\bf14} (3) (2017), 1730001 (27 pages).
\bibitem{Mantica5}C.A. Mantica, U.C. De, Y.J. Suh and L.G. Molinari, {\it Perfect fluid spacetimes with harmonic generalized curvature tensor}, Osaka J.Math., {\bf 56}, (2019), 173-182.

\bibitem{mrs}Mishra, R. S.,{\it Structures on Differentiable manifold and their applications},
           Chandrama Prakasana, Allahabad, 1984.

\bibitem{NO03}Nojiri, S. and Odintsov, S.D., Modified gravity with negative and positive powers of curvature: Unification of inflation and cosmic acceleration, Phys. Rev. D, {\bf68}(2003), Article ID 123512.

\bibitem{O83}O’Neill, B., Semi-Riemannian Geometry with Applications to the Relativity, Academic Press, New York-London, 1983.
\bibitem{ord}Ordines, T.M., Carson, E.D., {\it Limits on f(R, T)-gravity from Earth’s atmosphere}, Phys. Rev. D. {\bf 99} (2019), 104052.
\bibitem{PB06}Perez Bergliaffa, S.E., Constraining $f\left(R\right)$ theories with the energy conditions, Phys. Lett. B, {\bf642}(2006), 311--314.

\bibitem{RBB92}Raychaudhuri, A.K., Banerji, S. and Banerjee, A., General relativity, astrophysics, and cosmology, Springer-Verlag New York, Inc. 1992.
\bibitem{sin} Singh, V., Singh, C.P., {\it Modified f(R, T) gravity theory and scalar field cosmology}, Astrophys. Space Sci. {\bf356} (2015), 153–162.
\bibitem{ste} Stephani, H., Kramer, D., Mac-Callum, M., Hoenselaers, C., and Herlt, E., {\it Exact Solutions of Einstein’s Field Equations}, Cambridge University Press, Cambridge 2009.

\bibitem{yb}Yano, K. and Bochner S., {\it Curvature and Betti numbers}, Annals of Math. Studies 32
           ( Princeton university Press) 1953.


\end{thebibliography}
\end{document}